\documentclass[amsmath,amssymb,aps,pra,10pt,twocolumn]{revtex4-1}

\usepackage{dcolumn}
\usepackage{graphicx}
\usepackage{color}

\begin{document}

\setlength{\tabcolsep}{1pt}

\title{Long range forces between polar alkali diatoms aligned by
external electric fields}

\author{Jason N. Byrd}
\author{John A. Montgomery, Jr.}
\author{Robin C\^{o}t\'{e}}
 \affiliation{Department of Physics, University of Connecticut, Storrs, CT 06269}

\begin{abstract}
Long range electrostatic, induction and dispersion coefficients including terms of order
$R^{-8}$ have been calculated by the sum over states method using time dependent
density functional theory.  We also computed electrostatic moments and static 
polarizabilities of the individual diatoms up to the octopole order using coupled
cluster and density functional theory. The laboratory-frame transformed electrostatic 
moments and van der Waals coefficients corresponding to the alignment of the diatomic 
molecules were found. We use this transformation to obtain the
coupling induced by an external DC electric field, and present values for all $XY$ 
combinations of like polar alkali diatomic molecules with atoms from Li to Cs.
Analytic solutions to the dressed-state laboratory-frame electrostatic moments
and long range intermolecular potentials are also given for the DC low-field limit.
\end{abstract}

\maketitle

\section{Introduction}

Advances in the formation of ultracold absolute ground state polar alkali
diatoms \cite{miranda2011,deiglmayr2011} open up avenues into many branches of
the physical sciences.  For chemical physics, applications of polar diatoms
range from precision spectroscopy \cite{demille2000,demille2008}, to the study
\cite{balakrishnan2001,quemener2009,ni2010,quemener2011} and control
\cite{krems2008} of cold chemical reactions.  Other areas of physics benefit
from the use of polar molecules, such as condensed matter physics
\cite{micheli2006}, with the search for novel quantum gases \cite{santos2000}
and phases \cite{recati2003}.  Furthermore, dipolar gases have been the subject
of much interest from the quantum information community
\cite{yelin2006,yelin2009,yelin2010}, and ideas of atom optics ({\it e.g.} using
evanecent wave mirrors \cite{segev1997}) have been generalized to polar
molecules \cite{shimshon2005a,shimshon2005b}.  The recent achievements in
molecular alignment and control \cite{holmegaard2009,nielsen2012} may also allow
to take advantage of the unique properties and possible control provided by
ultracold polar molecules.  In addition, there is growing interest in reactions
of alkali diatoms to form tetramer
structures\cite{byrd2010,zuchowski2010,byrd2012-a} with reasonable dipole
moments and rich molecular structures, which could offer good candidates for
quantum computing with dipoles \cite{wei2011}.  In each of these applications it
is crucial to accurately describe the inter-molecular interactions, themselves
dominated by their long range behavior \cite{weck2006} at the low temperatures
found in these systems.  Because of the weakness of the long range
intermolecular forces as compared to the chemical bond, and the range of nuclear
coordinates and phase space involved, it is advantageous to consider alternative
methods of modeling the intermolecular potential other than {\it ab initio}
quantum chemical calculations.

\begin{figure}[b]
\caption{\label{geomfig}(a) Schematic representation of an aligned diatomic
molecule.  Classically, the molecule precesses on a cone of angle $\theta$
about the electric field ${\bf F}$, with $\langle\cos\theta\rangle$ describing the 
average orientation of the molecule: its dipole moment ${\cal D}$ points towards 
${\bf F}$ for $\langle\cos\theta\rangle>0$, and in the opposite direction for 
$\langle\cos\theta\rangle>0$. The alignment, 
$\langle\cos^2\theta\rangle$, describes the tightness of the rotational cone.
(b) Lab-fixed frame molecular interaction geometry in the presence of an
external field, where $\theta_F$ is the angle between the field and
the vector ${\bf R}$ joining the two molecules.}
\resizebox{8.5cm}{!}{\includegraphics{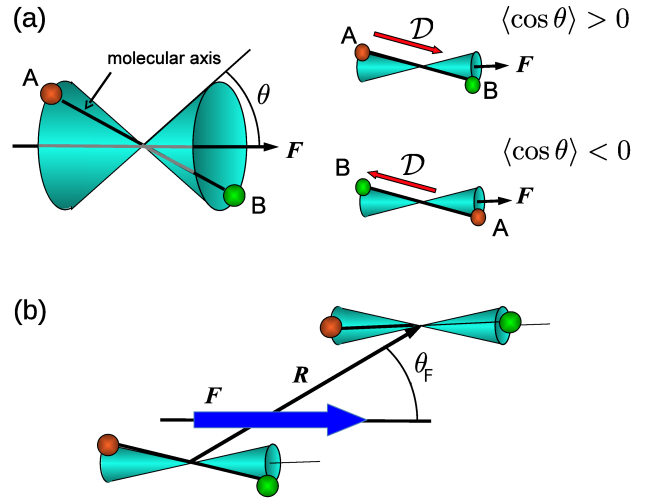}}
\end{figure}

A standard approach to describing the long range interaction potential between two
molecules, in the limit that the wavefunction overlap between the molecules is
negligible, is to expand the interaction energy into three distinct components,
\begin{equation}
E_{\rm int} = E_{\rm el} + E_{\rm ind} + E_{\rm disp}.
\end{equation}
Here $E_{\rm el}$, $E_{\rm ind}$ and $E_{\rm disp}$ are the permanent
electrostatic, induction (permanent-induced electrostatic) and dispersion
energies.  Each of these terms can be perturbatively expanded in an asymptotic
van der Waals series,
\begin{equation}
E_{\rm LR} = \sum_{n} C_n R^{-n}.
\end{equation}
The coefficients $C_n$ are in general angular dependent, and can be computed in
several ways.  In this work we expand the intermolecular electronic interaction
operator in a multipole expansion \cite{buckingham1967}, and then use first- and
second-order perturbation theory to calculate the van der Waals coefficients.
Several papers have discussed the isotropic $R^{-6}$ interactions of homonuclear
alkali diatoms using both the London approximation \cite{zemke2010} and time
dependent density functional theory (TD-DFT) \cite{banerjee2007,banerjee2009}.
The isotropic and anisotropic contributions have been investigated using
configuration interaction \cite{spelsberg93} and TD-DFT \cite{byrd2011} to
compute van der Waals coefficients through $R^{-8}$.  However, systematic
research on the heteronuclear alkali diatoms is limited to the $R^{-6}$
isotropic van der Waals coefficients for the LiX (X=Na,K,Rb,Cs) species
\cite{quemener2011}.  To date, the only heteronuclear anisotropic van der Waals
coefficients available in the literature are for KRb and RbCs
\cite{kotochigova2010} and limited to $R^{-6}$ dispersion forces.  In this paper
we present a systematic study of the isotropic and anisotropic van der Waals
interactions through order $R^{-8}$ of the heteronuclear alkali rigid-rotor
diatoms in their absolute ground state as a continuation of our work on the
homonuclear species \cite{byrd2011}.  Also included is the transformation of the
long range interaction potential from the molecule-fixed (MF) frame to the
laboratory-fixed (LF) frame for use in molecular alignment computations.  After
a brief description of dressed state diatomic molecules in Sec. \ref{align}, we
review the sum over states method of calculating van der Waals coefficients in
Sec. \ref{vdwsec}.  In Sec. \ref{dssec} the transformation to, and matrix
elements of, the lab-fixed frame van der Waals interaction potential are
described.  Analytic expressions of the low-field field coupled electrostatic
moments and van der Waals coefficients are also provided.  The {\it ab initio}
methodology is outlined in Sec. \ref{abinitio} and we conclude in Sec.
\ref{ressec} with a discussion of our numerical results.

\begin{table}[b]
\caption{\label{statictable}Center of mass multipole electrostatic moments, 
$\langle Q_{\ell 0}\rangle$, of all the ground 
state heteronuclear alkali diatoms through cesium evaluated
at the equilibrium bond length\footnote{$r_e$ values are taken from experimental
results where available, see Deiglmayr {\it et al.} \cite{deiglmayr2008} and
references therein.} $r_e$.  The variable $R_q$ denotes the distance where the
$R^{-5}$ electrostatic term overcomes the dipole-dipole $R^{-3}$ contribution.
All values are presented in atomic units.}
\begin{ruledtabular}
\begin{tabular}{llddddd}
    System & 
	Method &
    \multicolumn{1}{c}{$r_e$\footnotemark[1]} & 
    \multicolumn{1}{c}{$\langle Q_{10}\rangle$} &
    \multicolumn{1}{c}{$\langle Q_{20}\rangle$} &
    \multicolumn{1}{c}{$\langle Q_{30}\rangle$} &
    \multicolumn{1}{c}{$R_q$} \\
\hline
LiNa & CCSD(T) & 5.45 & 0.20 & 10.07 & -47.33 & 95 \\
     & VCI\footnote{Ref. \cite{aymar2005}.} & 5.43 & 0.22 & & & \\
	 & CCSDT\footnote{Ref. \cite{quemener2011}.} & 5.45 & 0.21 \\
	 \\
LiK  & CCSD(T) & 6.27 & 1.39 & 6.07 & -59.99 & 15 \\
     & VCI\footnotemark[2] & 6.21 & 1.39 & & & \\
	 & CCSDT\footnotemark[3] & 6.27 & 1.38 \\
	 \\
LiRb & CCSD(T) & 6.50 & 1.63 & 2.76 & -62.41 & 13 \\
     & VCI\footnotemark[2]  & 6.48 & 1.63 & & & \\
	 & CCSDT\footnotemark[3] & 6.50 & 1.59 \\
	 \\
LiCs & CCSD(T) & 6.93 & 2.15 & -2.29 & -49.88 & 10 \\
     & VCI\footnotemark[2]  & 6.82 & 2.17 & & & \\
	 & CCSDT\footnotemark[3] & 6.93 & 2.11 \\
	 \\
NaK  & CCSD(T) & 6.61 & 1.12 & 10.56 & -26.54 & 19 \\
     & CCSD(T)\footnote{Ref. \cite{zemke2010}.} & 6.592 & 1.156 & 10.60 & & \\
     & VCI\footnotemark[2]  & 6.49 & 1.09 & & & \\
	 \\
NaRb & CCSD(T) & 6.88 & 1.35 & 6.94 & -56.00 & 16 \\
     & VCI\footnotemark[2]  & 6.84 & 1.30 & & & \\
	 \\
NaCs & CCSD(T) & 7.27 & 1.85 & 2.49 & -60.45 & 12 \\
     & VCI\footnotemark[2]  & 7.20 & 1.83 & & & \\
	 \\
KRb  & CCSD(T) & 7.69 & 0.25 & 15.14 & -69.09 & 109 \\
     & VCI\footnotemark[2]  & 7.64 & 0.23  & & & \\
 & rel\footnote{Ref. \cite{kotochigova2003} performed a
 four component Dirac-Fock valence bond calculation in calculating the dipole
 moment.} & 7.7 & 0.30 & & & \\
 \\
KCs  & CCSD(T) & 8.10 & 0.75 & 13.00 & -105.70 & 38 \\
     & VCI\footnotemark[2]  & 8.02 & 0.76 & & & \\
	 \\
RbCs & CCSD(T) & 8.37 & 0.49 & 15.88 & -50.28 & 60 \\
     & VCI\footnotemark[2]  & 8.30 & 0.40 & & & 
\end{tabular}
\end{ruledtabular}
\end{table}

\section{\label{align}Dressed state diatomic molecules}

\begin{table*}[t]
\caption{\label{polartable}Multipole static polarizabilities, $\alpha_{\ell\ell' m},$ and isotropic van
der Waals dispersion coefficients, $W^{(2,{\bf DIS})}_{n000},$ up to order $n=8$
of all the ground state alkali diatoms through cesium evaluated at the
equilibrium bond lengths $r_e$ listed in Table \ref{statictable}.  
All values are presented in atomic units.}
\begin{ruledtabular}
\begin{tabular}{lldddddddd}
    System & 
	Method &
	\multicolumn{1}{c}{$\alpha_{110}$\footnote{Note that the parallel and
	perpendicular static dipole polarizabilities, $\alpha_\|$ and $\alpha_\bot$,
	correspond to $\ell\ell' m=110$ and $111$ respectively.}
	}  & 
    \multicolumn{1}{c}{$\alpha_{111}$\footnotemark[1]}  & 
    \multicolumn{1}{c}{$\bar{\alpha}$\footnote{$\bar{\alpha}=\frac{1}{3}(\alpha_\|+2\alpha_\bot)$
	is the average static dipole polarizability.}}  & 
    \multicolumn{1}{c}{$\alpha_{220}$}  & 
    \multicolumn{1}{c}{$\alpha_{221}$}  & 
    \multicolumn{1}{c}{$\alpha_{222}$}  & 
    \multicolumn{1}{c}{$W^{(2,{\bf DIS})}_{6000}$} &
    \multicolumn{1}{c}{$W^{(2,{\bf DIS})}_{8000}$} \\
\hline
LiNa & PBE0 & 300.0 & 185.5 & 223.7 & 9418.9 & 7035.6 & 3356.2 & 3.279[3] & 4.982[5] \\
     & VCI\footnote{Ref. \cite{deiglmayr2008}.} & 347.6 & 181.8 & 237.0 \\
	 & CCSDT\footnote{Ref. \cite{quemener2011}.} &  &  & 237.8 &  &  &  &
	 3.673[3]\footnote{Ref. \cite{quemener2011} evaluated using CCSD and the
	 Tang-Slater-Kirkwood formula \cite{tang1969}.} \\
	 \\
LiK  & PBE0 & 455.1 & 261.8 & 326.3 & 24164.4 & 15899.8 & 5939.6 & 5.982[3] & 1.378[6] \\
     & VCI\footnotemark[4]      & 489.7 & 236.2 & 320.7 \\
	 & CCSDT\footnotemark[5]     &  &  & 324.9  &  &  &  & 6.269[3]\footnotemark[6] \\
	 \\
LiRb & PBE0 & 445.5 & 256.1 & 319.2 & 27815.3 & 18110.7 & 6359.2 & 6.193[3] & 1.583[6] \\
     & VCI\footnotemark[4]      & 524.3 & 246.5 & 339.1 \\
	 & CCSDT\footnotemark[5]     &  &  & 346.2 &  &  &  & 6.323[3]\footnotemark[6] \\
	 \\
LiCs & PBE0 & 525.2 & 289.1 & 367.8 & 38723.9 & 24996.3 & 7935.8 & 7.700[3] & 2.297[6] \\
     & VCI\footnotemark[4]      & 597.0 & 262.5 & 374.0 \\
	 & CCSDT\footnotemark[5]     &  &  & 386.7 &  &  &  & 7.712[3]\footnotemark[6] \\
	 \\
NaK  & PBE0 & 472.7 & 280.6 & 344.6 & 16572.0 & 13035.0 & 6739.5 & 6.818[3] & 1.268[6] \\
     & VCI\footnotemark[4]      & 529.2 & 262.3 & 351.3 \\
	 & CCSD(T)\footnote{Ref. \cite{zemke2010}.} &  &  & 363.8 &  &  &  &
	 6.493[3]\footnote{Ref. \cite{zemke2010} evaluated using the London
	 formula.} \\
	 \\
NaRb & PBE0 & 504.6 & 285.3 & 358.4 & 25217.0 & 17771.7 & 7547.5 & 7.688[3] & 1.790[6] \\
     & VCI\footnotemark[4]      & 572.0 & 280.3 & 377.5 \\
	 \\
NaCs & PBE0 & 587.3 & 323.2 & 411.2 & 37633.3 & 25245.7 & 9444.6 & 9.453[3] & 2.641[6] \\
     & VCI\footnotemark[4]      & 670.7 & 304.2 & 426.4 \\
	 \\
KRb  & PBE0 & 729.6 & 420.9 & 523.8 & 36974.1 & 27588.8 & 13100.9 & 1.349[4] & 3.385[6] \\
     & VCI\footnotemark[4]      & 748.7 & 382.9 & 504.8 \\
	 \\
KCs  & PBE0 & 836.7 & 468.6 & 591.3 & 56372.9 & 38791.7 & 16262.9 & 1.657[4] & 5.038[6] \\
     & VCI\footnotemark[4]      & 822.3 & 425.62& 571.1 \\
	 \\
RbCs & PBE0 & 901.0 & 502.0 & 635.0 & 48325.3 & 36401.8 & 18619.8 & 1.884[4] & 5.188[6] \\
     & VCI\footnotemark[4]      & 904.0 & 492.3 & 602.8
\end{tabular}
\end{ruledtabular}
\end{table*}

The orientation and alignment ($\langle\cos\theta\rangle$ and
$\langle\cos^2\theta\rangle$ \footnote{The orientation and alignment of a dipole
in an external field is given by the dipole-vector and field-vector correlation
distribution.  Expanding the distribution in terms of Legendre polynomials, the
first odd term is $\cos\theta$ corresponding to the {\it orientation cosine} and
the first even term is $\cos^2\theta$ corresponding to the {\it alignment
cosine}.} respectively as illustrated in Fig.  \ref{geomfig}(a)) of polar
molecules can be achieved through several mechanisms, the most direct of which
is the coupling of rotational states by a polarizing external DC electric field,
${\bf F}$.  Increasing the strength, $F$, of the external electric field
increases the number of rotational states coupled, thus tightening the
orientation of the molecule in a cone of angle $\theta$ about the orientation of
the field.  To account for this rotational coupling adiabatically, we expand the
dressed state rotational wave function of the molecule as a superposition of
field-free symmetric top states 
\begin{equation}\label{coupledj}
|\tilde{J}\tilde{M}\Omega \rangle=\sum_{J,M}
a_{M}^{J}|JM\Omega \rangle,
\end{equation}
where the symmetric top states are given in term of 
Wigner rotation matrices $D^{J}_{-M-\Omega}(\alpha,\beta,\gamma)$
\cite{edmonds57}
\begin{equation}
|JM\Omega \rangle = 
(-1)^{M-\Omega}
\left(\frac{2J+1}{8\pi^2}\right)
D^{J}_{-M-\Omega}(\alpha,\beta,\gamma),
\end{equation}
where $(\alpha,\beta,\gamma)$ are the Euler angles of the molecule, and $J$ the
total angular momentum quantum number with projections $M$ in the laboratory
frame and $\Omega$ onto the molecular axis.  The expansion coefficients
$a_{M}^{J}$ dictate the levels of mixing between the different rotational
states, and can be solved for by diagonalizing
$\langle\tilde{J}'\tilde{M}'\Omega'|H|\tilde{J}\tilde{M}\Omega\rangle$.  Here
$H$ is the symmetric top and dipole-field Hamiltonian
\begin{equation}\label{rothamiltonian}
H = B({\bf J}^2+J^2_z) - {\cal D} F \cos\theta,
\end{equation}
where ${\bf J}$ is the angular momentum operator, $J_z$ is the angular momentum
projection on the $z$ axis, $B$ is the molecular rotational constant, $\cal D$
is the dipole moment of the molecule, and $\theta$ is the angle between the
external electric field of magnitude $F$ and the molecular axis. 
The coefficients $a_{M}^{J}(F)$ then depend on the strength $F$ of the field.
While theoretically simple, this process can become experimentally challenging for
molecules with small dipole moments or rotational constants due to the large
external fields required for strong alignment.

An alternative to simply increasing the static field magnitude is to add a separate
polarizing laser field \cite{hartelt2008} that directly couples the rotational
states of the molecule.  However, to achieve both alignment and
orientation control, time-dependent nonadiabatic effects are introduced into the
dressed state wavefunction \cite{nielsen2012}.  For the purposes of this work
the investigation and inclusion of these nonadiabatic effects are unimportant as
only the final dressed state is of interest.  As such we present our alignment
in terms of an applied external static field and, where practical, the number of
strongly coupled rotational states.

\begin{table}[t]
\caption{\label{livdwtable}LiX (X=Na,K,Rb,Cs) calculated CCSD(T) electrostatic
and TD-DFT dispersion$+$induction van der Waals coefficients, $W^{(1,2)}_{n L_1
L_2 M},$ for unique combinations of $ L_1 L_2 M$.  All values are presented in
atomic units and calculated at the equilibrium bond length $r_e$ listed in Table
\ref{statictable}, and [n] denotes $\times 10^n$.} 
\begin{ruledtabular}
\begin{tabular}{crrrr}
$n~L_1~L_2~M$ & 
    \multicolumn{1}{c}{LiNa} & 
    \multicolumn{1}{c}{LiK}  & 
    \multicolumn{1}{c}{LiRb} & 
    \multicolumn{1}{c}{LiCs} \\
\hline
\multicolumn{4}{l}{Electrostatic: $W^{(1)}_{nL_1L_2M}$} \\
3110 & -7.076[0] & -3.799[0] & -5.170[0] & -9.094[0]  \\
3111 & 0.038[0] & 1.900[0] & 2.585[0] & 4.547[0]  \\
\\
4210 & -5.893[0] & -2.547[1] & -1.431[1] & 1.324[1]  \\
4211 & 1.965[0] & 8.489[0] & 4.771[0] & -4.412[0]  \\
\\
5220 & 6.086[2] & 2.276[2] & 5.284[1] & 2.568[1]  \\
5221 & -1.353[2] & -5.058[1] & -1.174[1] & -5.707[0]  \\
5222 & 8.453[0] & 3.161[0] & 0.734[0] & 0.357[0]  \\
5310 & 3.674[1] & 3.306[2] & 2.887[2] & 3.942[2]  \\
5311 & -9.186[0] & -8.266[1] & -7.218[1] & -9.855[1]  \\
\multicolumn{4}{l}{Dispersion$+$Induction:  $W^{(2)}_{nL_1L_2M}$} \\
6000 & 3.289[3] & 7.243[3] & 7.254[3] & 1.062[4]  \\
6200 & 4.036[2] & 3.479[2] & 4.874[2] & 2.739[1]  \\
6220 & 1.768[2] & 1.180[3] & 1.094[3] & 2.567[3]  \\
6221 & -3.929[1] & -2.621[2] & -2.430[2] & -5.705[2]  \\
6222 & 4.911[0] & 3.277[1] & 3.038[1] & 7.131[1]  \\
\\
7100 & 1.075[3] & -4.898[2] & -5.906[3] & -3.740[4]  \\
7210 & 1.460[2] & -7.641[3] & -8.641[3] & -3.110[4]  \\
7211 & -2.433[1] & 1.273[3] & 1.440[3] & 5.184[3]  \\
7300 & 7.411[2] & 4.567[3] & 1.057[3] & -9.020[3]  \\
7320 & 4.259[2] & -7.013[2] & -2.682[3] & -1.687[4]  \\
7321 & -7.099[1] & 1.169[2] & 4.470[2] & 2.812[3]  \\
7322 & 5.070[0] & -8.349[0] & -3.193[1] & -2.008[2]  \\
\\
8000 & 5.586[5] & 1.539[6] & 1.715[6] & 2.722[6]  \\
8200 & 3.552[5] & 1.270[6] & 1.534[6] & 2.793[6]  \\
8220 & 9.460[4] & 5.702[5] & 6.695[5] & 1.636[6]  \\
8221 & -1.406[4] & -8.078[4] & -9.379[4] & -2.272[5]  \\
8222 & 1.234[3] & 4.607[3] & 4.642[3] & 9.978[3]  \\
8400 & 3.896[4] & 4.574[4] & 6.320[4] & 1.371[5]  \\
8420 & 2.706[4] & 7.284[4] & 8.377[4] & 2.485[5]  \\
8421 & -3.666[3] & -9.822[3] & -1.124[4] & -3.320[4]  \\
8422 & 2.001[2] & 5.287[2] & 5.965[2] & 1.740[3]  
\end{tabular}
\end{ruledtabular}
\end{table}

\section{\label{vdwsec}Anisotropic long range interactions}

Given a linear molecule in the Born-Oppenheimer approximation (no nuclear
motion), at any given configuration the orientation of each molecule can be
described by the vector $\hat{r}_i = (\theta_i,\phi_i)$ with the relative
position between the molecular center of mass defined as ${\bf
R}=(R,\theta,\phi)$.  Here $\theta_i$ is the projection angle of $\hat{r}_i$ on
${\bf R}$, $\phi_i$ is the projection angle of $\hat{r}_i$ on the $x$ axis and
$(R,\theta,\phi)$ are the spherical vector components of $\bf R$.
Due to the rotational invariance of the interaction energy
between two molecules, it can be separated into a series of radial and angular
basis functions
\begin{equation}\label{gvdw}
E_{\rm int} (\hat{r}_1,\hat{r}_2,{\bf R}) = \!\! \sum_{L_1,L_2,L} \!\!
E_{L_1 L_2 L}(R) A_{L_1 L_2 L}(\hat{r}_1,\hat{r}_2,\hat{R}).
\end{equation}
Here $E_{L_1 L_2 L}(R)$ are purely radial functions for a rigid-rotor and $A_{L_1 L_2
L}(\hat{r}_1,\hat{r}_2,\hat{R})$ is an angular basis which, when ${\bf R}$ is
oriented along the $z$ axis, can be expressed as \cite{mulder1979}
\begin{multline}\label{asimple}
A_{L_1 L_2 L}(\hat{r}_1,\hat{r}_2,\hat{R}) = \sum_{M=0}^{min(L_1,L_2)}
\eta^{M}_{L_1 L_2 L} \\ \times
P^{M}_{L_1}(\cos\theta_1) P^{M}_{L_2}(\cos\theta_2)\cos [M(\phi_1-\phi_2)],
\end{multline}
where
\begin{multline}
\eta^M_{L_1 L_2 L} = (-1)^M (2-\delta_{M,0}) (L_1 M;L_2 -M|L 0) \\ \times
\left[\frac{(L_1-M)!  (L_2-M)!} {(L_1+M)!(L_2+M)!}\right]^{1/2},
\end{multline}
$(L_1 M;L_2 -M|L 0)$ is a Clebsch-Gordon coefficient, and $P_L^M(\cos\theta)$
is an associated Legendre polynomial.  The radial functions $E_{L_1 L_2 L}(R)$
can be evaluated using first- and second-order perturbation theory by expanding
in terms of the electronic multipole operators
$Q_{\ell m}=\sum_i z_i r^\ell_i C_{\ell m}(\hat{r}_i)$,
where the sum is over all charges, $z_i$ is the charge at each $i$'th center,
$r^\ell_i$ is the distance from each $i$'th charge to the center of mass and
$C_{\ell m}(\hat{r}_i)$ is a Racah spherical harmonic \cite{edmonds57}.

Following the standard approach \cite{buckingham1967,avoird1980,byrd2011}, the
first- and second-order interaction energy for two linear molecules can be
expressed as
\begin{multline}\label{wsimple}
E_{\rm int}(R,\theta_1,\theta_2,\phi) = 
\sum_{n,L_1,L_2,M} \frac{(W^{(1)}_{n L_1 L_2 M}-W^{(2)}_{n L_1 L_2 M})}{R^n} \\ 
\times 
P^{M}_{L_1}(\cos\theta_1) P^{M}_{L_2}(\cos\theta_2)\cos [ M\phi ] ,
\end{multline}
where $\phi\equiv \phi_1-\phi_2,$
\begin{multline}\label{w1}
W^{(1)}_{nL_1L_2M} =
(-1)^{L_1+M}(2-\delta_{M,0})\frac{(L_1+L_2)!}{(L_1+M)!(L_2+M)!} \\ \times
\langle 0_1|Q_{L_1 0}|0_1\rangle \langle 0_2|Q_{L_2 0}|0_2\rangle
\end{multline}
is the first-order electrostatic contribution, where $|0_i\rangle$
is the electronic ground state of molecule $i$. In Eq.(\ref{wsimple}),
\begin{equation}
W^{(2)}_{n L_1 L_2 M} = W^{(2,{\bf DIS})}_{n L_1 L_2 M}+W^{(2,{\bf IND})}_{n L_1 L_2 M}
\end{equation}
contains the second-order contributions from dispersion,
\begin{multline}\label{w2dis}
W^{(2,{\bf DIS})}_{n L_1 L_2 M}(R) = 
\sum_{\substack{\ell_1,\ell'_1\\ \ell_2,\ell'_2}} 
\zeta^{\ell_1 \ell'_1;\ell_2 \ell'_2}_{L_1L_2M} 
\delta_{\ell_1+ \ell'_1+ \ell_2+ \ell'_2+2,n}\\ \times
\sum_{\substack{ k_1\ne 0\\ k_2\ne 0 }} 
\frac{T^{0_1 k_1}_{\ell_1 \ell'_1 L_1} T^{0_2 k_2}_{\ell_2 \ell'_2 L_2}}
{\epsilon_{k_1}-\epsilon_{0_1}+ \epsilon_{k_2}-\epsilon_{0_2}},
\end{multline}
and induction,
\begin{multline}\label{w2ind}
W^{(2,{\bf IND})}_{n L_1 L_2 M}(R) =
\sum_{\substack{\ell_1,\ell'_1\\ \ell_2,\ell'_2}} 
\zeta^{\ell_1 \ell'_1; \ell_2 \ell'_2}_{L_1L_2M} 
\delta_{\ell_1+ \ell'_1+ \ell_2+ \ell'_2+2,n}\\ \times
\left(
 T^{0_1 k_1}_{\ell_1 \ell'_1 L_1}
\sum_{k_2\ne 0} 
\frac{T^{0_2 k_2}_{\ell_2 \ell'_2 L_2}}{\epsilon_{k_2}-\epsilon_{0_2}}
+(1\rightleftharpoons 2)
\right).
\end{multline}
The scalar coupling coefficient $\zeta^{\ell_1 \ell'_1;\ell_2 \ell'_2}_{L_1L_2M}$ 
is given \cite{mulder1979} as
\begin{multline}\label{zeta}
\zeta^{\ell_1 \ell'_1;\ell_2 \ell'_2}_{L_1L_2M} =
(-1)^{\ell_2+\ell'_2}
((2L_1+1)!  (2L_2+1)!)^{1/2}\\
\times \left[\frac{ (2\ell_1+2\ell_2+1)!  (2\ell'_1+2\ell'_2+1)! }
        { (2\ell_1)!  (2\ell'_1)!  (2\ell_2)!  (2\ell'_2)!}\right]^{1/2}
\sum_L \eta^{M}_{L_1,L_2,L} \\
\times (\ell_1+\ell_2 0;\ell'_1+\ell'_2 0|L 0)
\begin{Bmatrix}
\ell_1 & \ell'_1 & L_1 \\
\ell_2 & \ell'_2 & L_2 \\
\ell_1+\ell_2 & \ell'_1+\ell'_2 & L
\end{Bmatrix},
\end{multline}
the symbol between curly brackets being a Wigner 9-j symbol \cite{edmonds57},
and $T^{0_i k_i}_{\ell_i \ell'_i L_i}$ is the coupled monomer multipole transition
moment defined as
\begin{equation}\label{coupledtm}
T^{0_i k_i}_{\ell_i \ell'_i L_i} = \sum_m \langle 0_i|Q_{\ell_i m}|k_i\rangle \langle
k_i|Q_{\ell'_i -m}|0_i\rangle (\ell_i m;\ell'_i -m|L_i 0) ,
\end{equation}
where the indices $k_i$ go over ground and excited states of the $i$'th
molecules electronic wavefunction $|k_i\rangle$ \footnote{In principle the
summation over $k_i$ in Eqs. \ref{w2dis} and \ref{w2ind} involve an integral-sum
over continuum states.  In practice we truncate the summation to include all
single excitations within the electronic Hilbert space.}.  It is convenient,
when discussing molecular properties, to work with the uncoupled dynamic
multipole polarizability:
\begin{equation}\label{ualphaw}
\alpha_{\ell \ell' m}(\omega) = \sum_{k\ne 0}
\frac{(\epsilon_{k}-\epsilon_{0}) 
\langle 0_i|Q_{\ell m}|k_i\rangle \langle k_i|Q_{\ell' -m}|0_i\rangle}
{(\epsilon_{k}-\epsilon_{0})^2 - \omega^2}.
\end{equation}
The zero frequency limit of Eq.(\ref{ualphaw}) represents the static multipole polarizability.  

\section{\label{dssec}Dressed-state van der Waals interaction}

\subsection{General Expressions}

\begin{table*}[floatfix]
\caption{\label{othervdwtable} XY (X,Y=Na,K,Rb,Cs) calculated CCSD(T) electrostatic
and TD-DFT dispersion$+$induction van der Waals coefficients, $W^{(1,2)}_{n L_1
L_2 M},$ for unique combinations of $ L_1 L_2 M$.  All values are presented in
atomic units and calculated at the equilibrium bond length $r_e$ listed in Table
\ref{statictable}, and [n] denotes $\times 10^n$.} 
\begin{ruledtabular}
\begin{tabular}{cdddddd}
$n~L_1~L_2~M$ & 
    \multicolumn{1}{c}{NaK}  & 
    \multicolumn{1}{c}{NaRb} & 
    \multicolumn{1}{c}{NaCs} & 
    \multicolumn{1}{c}{KRb}  & 
    \multicolumn{1}{c}{KCs}  & 
    \multicolumn{1}{c}{RbCs} \\
\hline
\multicolumn{4}{l}{Electrostatic: $W^{(1)}_{nL_1L_2M}$} \\
3110 & -2.470[0] & -3.651[0] & -6.813[0] & -0.125[0] & -1.164[0] & -0.432[0] \\
3111 &  1.235[0] &  1.826[0] &  3.406[0] &  0.063[0] &  0.582[0] & 0.216[0] \\
\\
4210 & -3.528[1] & -2.818[1] & -1.377[1] & -1.134[1] & -2.960[1] & -2.221[1] \\
4211 &  1.176[1] &  9.393[0] &  4.590[0] &  3.785[0] &  9.867[0] & 7.403[0] \\
\\
5220 & 6.717[2]  & 2.900[2]  & 3.710[1]  & 1.375[3]  & 1.004[3]  & 1.524[3] \\
5221 & -1.493[2] & -6.444[1] & -8.245[0] & -3.056[2] & -2.231[2] & -3.386[2] \\
5222 & 9.329[0]  & 4.028[0]  & 0.515[0]  & 1.910[1]  & 1.395[1]  & 2.116[1] \\
5310 & 1.175[2]  & 7.346[1]  & 5.962[1]  & 6.909[1]  & 1.955[2]  & 1.106[1] \\
5311 & -2.937[1] & -1.836[1] & -1.490[1] & -1.727[1] & -4.887[1] & -2.764[0] \\
\multicolumn{4}{l}{Dispersion$+$Induction: $W^{(2)}_{nL_1L_2M}$} \\
6000 & 7.777[3] & 8.680[3] & 1.233[4] & 1.354[4] & 1.726[4] & 1.921[4] \\
6200 & 5.519[2] & 7.837[2] & 3.327[2] & 2.001[3] & 2.375[3] & 2.909[3] \\
6220 & 9.762[2] & 1.223[3] & 2.694[3] & 1.028[3] & 1.826[3] & 1.857[3] \\
6221 & -2.169[2] & -2.717[2] & -5.986[2] & -2.284[2] & -4.059[2] & -4.127[2] \\
6222 & 2.712[1] & 3.397[1] & 7.482[1] & 2.855[1] & 5.073[1] & 5.159[1] \\
\\
7100 & 1.157[4] & 2.268[3] & -2.114[4] & 6.069[3] & 1.457[4] & 1.755[4] \\
7210 & -1.657[1] & -4.434[3] & -2.312[4] & 9.929[2] & -8.871[2] & 2.795[3] \\
7211 & 2.762[0] & 7.391[2] & 3.853[3] & -1.655[2] & 1.479[2] & -4.658[2] \\
7300 & 9.136[3] & 4.795[3] & -1.280[3] & 4.176[3] & 1.234[4] & 1.231[4] \\
7320 & 4.931[3] & 1.142[3] & -9.091[3] & 2.772[3] & 7.134[3] & 8.669[3] \\
7321 & -8.219[2] & -1.903[2] & 1.515[3] & -4.621[2] & -1.189[3] & -1.445[3] \\
7322 & 5.870[1] & 1.360[1] & -1.082[2] & 3.300[1] & 8.492[1] & 1.032[2] \\
\\
8000 & 1.444[6] & 1.928[6] & 3.016[6] & 3.734[6] & 5.391[6] & 5.667[6] \\
8200 & 8.920[5] & 1.503[6] & 2.766[6] & 2.606[6] & 4.295[6] & 3.923[6] \\
8220 & 3.296[5] & 6.003[5] & 1.466[6] & 7.941[5] & 1.494[6] & 1.296[6] \\
8221 & -4.768[4] & -8.470[4] & -2.042[5] & -1.162[5] & -2.138[5] & -1.878[5] \\
8222 & 3.368[3] & 4.596[3] & 9.349[3] & 9.098[3] & 1.356[4] & 1.351[4] \\
8400 & 6.476[4] & 6.275[4] & 1.107[5] & 2.931[5] & 3.258[5] & 4.238[5] \\
8420 & 5.979[4] & 7.214[4] & 1.962[5] & 2.200[5] & 2.678[5] & 3.389[5] \\
8421 & -8.081[3] & -9.702[3] & -2.625[4] & -2.970[4] & -3.612[4] & -4.566[4] \\
8422 & 4.380[2] & 5.183[2] & 1.380[3] & 1.603[3] & 1.945[3] & 2.453[3]
\end{tabular}
\end{ruledtabular}
\end{table*}

To consider the interactions between rigid-rotor linear molecules dressed by an
external electric field it is necessary to first transform the van der Waals
interaction energy from the molecule-fixed frame (MF) to the lab-fixed frame
(LF).  The lab-fixed frame van der Waals interaction can be generally expressed
by referring to Eq.(\ref{gvdw}) and removing the constraint on
Eq.(\ref{asimple}) which specified that ${\bf R}$ is aligned to the $z$-axis.
The angular basis can then generally be expressed
\cite{avoird1980} as
\begin{multline}\label{ageneral}
A_{L_1 L_2 L}(\hat{R},\hat{r}_1,\hat{r}_2) =
\sum_{ m_{L_1},m_{L_2},m_L}
\left(\begin{matrix}
L_1 & L_2 & L \\
m_{L_1} & m_{L_2} & m_L
\end{matrix}\right) \\ \times
Y_{L_1 m_{L_1}}(\hat{r}_1)
Y_{L_2 m_{L_2}}(\hat{r}_2)
Y_{L m_{L}}(\hat{R}),
\end{multline}
where $Y_{\ell m_\ell}(\hat{r})$ is a spherical harmonic and $(:::)$ is a Wigner 3-j
symbol \cite{edmonds57}.  Because of the change in
angular basis, it is necessary to recouple the radial $W^{(1,2)}_{nL_1 L_2 M}$
functions.  This can be done readily by integrating Eq.(\ref{wsimple}) over the
angular phase space:
\begin{multline}\label{fieldoverlap}
E^{\rm LF}_{L_1 L_2 L}(R) = \frac{1}{\sqrt{8\pi}}
\int^{2\pi}_0 d\phi \int^{\pi}_0 d\theta_1 \int^{\pi}_0 d\theta_2
\sin(\theta_1)\sin(\theta_2) \\ \times
A'_{L_1 L_2 L}(\theta_1 \theta_2 \phi)
E^{\rm MF}_{\rm int} (R,\theta_1,\theta_2,\phi),
\end{multline}
where $A'_{L_1,L_2,L}(\theta_1,\theta_2,\phi)$ is Eq.(\ref{ageneral}) 
projected onto the molecule-fixed frame and is given by
\cite{tscherbul2009} 
\begin{multline}
A'_{L_1 L_2 L}(\theta_1,\theta_2,\phi) =
\left(\frac{2L+1}{2\pi}\right)^{1/2} \\ \times
\sum_{m=0}^{\min(L_1,L_2)}
(-1)^m (2 - \delta_{m,0})
\left(\begin{matrix}
L_1 & L_2 & L \\
m         & -m        & 0
\end{matrix}\right)\\ \times
\Theta_{L_1 m}(\theta_1)\Theta_{L_2 m}(\theta_2) \cos [m\phi].
\end{multline}
where $\Theta_{l m}(\theta)$ are normalized associated Legendre polynomials.
The resulting integrand has the solution 
\begin{multline}\label{vint}
E^{\rm LF}_{L_1 L_2 L}(R) =  \sum_M (-1)^M 
\left(\begin{matrix}
L_1 & L_2 & L \\
M         & -M        & 0
\end{matrix}\right) \\ \times
\left(
\frac{(L_1+M)!(L_2+M)!}{(L_1-M)!(L_2-M)!}
\right)^{1/2} 
\left(
\frac{2L+1}{(2L_1 + 1)(2L_2 + 1)}
\right)^{1/2} \\ \times
\sum_n \frac{(W^{(1)}_{n L_1 L_2 M}-W^{(2)}_{n L_1 L_2 M})}{R^n}.
\end{multline}

The adiabatic dressed state basis for two molecules at large separation is given in terms
of the product of each molecule dressed rotational wave functions
\begin{equation}\label{dbasis}
\phi = |\tilde{J_1}\tilde{M_1}\Omega_1\rangle\otimes
|\tilde{J_2}\tilde{M_2}\Omega_2\rangle.
\end{equation}
The dressed state (DS) van der Waals interaction $E^{\rm DS}_{\rm int}({\bf R})$
is calculated from the matrix elements of Eq.(\ref{vint}) in the dressed state
basis, 
\begin{widetext}
\begin{multline}\label{transvdw}
E^{\rm DS}_{\rm int}({\bf R}) = 
\langle\tilde{J}'_2\tilde{M}'_2\Omega'_2|\langle\tilde{J}'_1\tilde{M}'_1\Omega'_1| 
E^{\rm LF}_{\rm int}(\hat{r}_1,\hat{r}_2,{\bf R})
|\tilde{J}_1\tilde{M}_1\Omega_1\rangle|\tilde{J}_2\tilde{M}_2\Omega_2\rangle
=
\sqrt{4\pi}
\sum_{\substack{ J_1,M_1 \\ J'_1,M'_1 }}
\sum_{\substack{ J_2,M_2 \\ J'_2,M'_2 }}
\sum_{\substack{L_1,m_{L_1}\\L_2,m_{L_2}}}
\delta_{\Omega_1,\Omega'_1}
\delta_{\Omega_2,\Omega'_2}
\\
(-1)^{M_1 - \Omega_1 + m_{L_1}}
(-1)^{M_2 - \Omega_2 + m_{L_2}}
\rho^{J`_1J_1}_{M'_1M_1}(F)
\rho^{J`_2J_2}_{M'_2M_2}(F)
\left[ 
(2J_1 + 1) (2J'_1 + 1) (2J_2 + 1) (2J'_2 + 1) (2L_1 + 1) (2L_2 + 1)
\right]^{1/2} \\ \times
\left(\begin{matrix}
J'_1 & L_1     &  J_1 \\
M'_1 & m_{L_1} & -M_1
\end{matrix}\right)
\left(\begin{matrix}
J'_1 & L_1   & J_1 \\
\Omega_1 & 0 & -\Omega_1
\end{matrix}\right)
\left(\begin{matrix}
J'_2 & L_2     &  J_2 \\
M'_2 & m_{L_2} & -M_2
\end{matrix}\right)
\left(\begin{matrix}
J'_2 & L_2   & J_2 \\
\Omega_2 & 0 & -\Omega_2
\end{matrix}\right)
\sum_{L,m_L} 
\left(\begin{matrix}
L_1     & L_2     & L \\
m_{L_1} & m_{L_2} & m_L
\end{matrix}\right)
Y_{L m_L}(\hat{R})
E^{\rm LF}_{L_1 L_2 L}(R),
\end{multline}
\end{widetext}
where
\begin{equation}\label{dens}
\rho^{J'_i J_i}_{M_i M'_i} (F) = a^{J'_i}_{M'_i}(F) a^{J_i}_{M_i}(F)
\end{equation}
is the coupled rotational state density of molecule $i$ and $Y_{Lm}(\hat{r})$ is
a spherical harmonic \cite{edmonds57}.  
In addition to the transformation of the van der Waals interaction energy as
given by Eq.(\ref{transvdw}), it is useful to have the dressed static moment,
$\langle Q^{DS}_{\ell 0}\rangle,$ of a given molecule.  For molecule $i$, this 
is readily obtained to be
\begin{multline}\label{dressedq}
\langle Q^{DS}_{\ell 0}\rangle = 
\langle\tilde{J}'_i\tilde{M}'_i\Omega'_i|\langle 0_i|
Q_{\ell 0}
|0_i\rangle|\tilde{J}_i\tilde{M}_i\Omega_i\rangle
= \\
\sum_{\substack{ J_i,J'_i \\ M_i }}
\delta_{\Omega_i,\Omega'_i}
\delta_{M_i,M'_i}
\rho^{J'_iJ_i}_{M_iM_i}
\left((2J_i+1)(2J_i'+1)\right)^{1/2} \\ \times
(-1)^{M_i - \Omega_i}
\left(\begin{matrix}
J'_i & \ell & J_i \\
M'_i & 0 & -M_i
\end{matrix}\right)
\left(\begin{matrix}
J'_i & \ell & J_i \\
\Omega_i & 0 & -\Omega_i
\end{matrix}\right)
\langle Q_{\ell 0}\rangle.
\end{multline}

\subsection{\label{lowfs}Low-Field Solution}

\begin{figure}[t]
\resizebox{8.5cm}{!}{\includegraphics{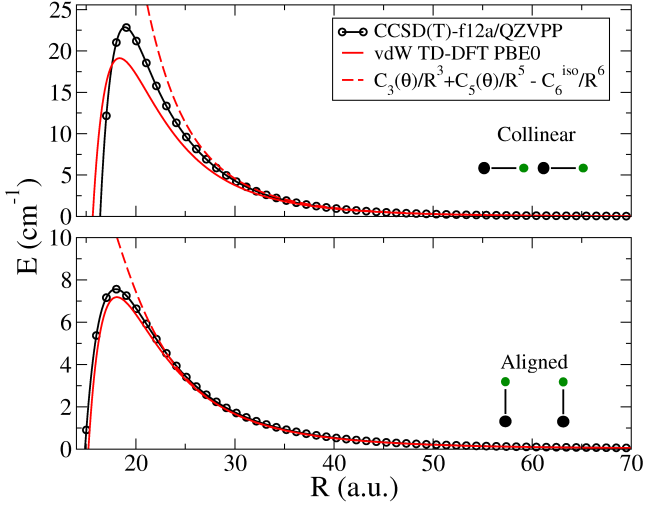}}
\caption{\label{abinitfig}Comparison between the TD-DFT van der Waals surface,
evaluated at both collinear ($\theta=0$) and aligned ($\theta=\pi/2$)
geometries, and both the electrostatic~$+$~isotropic $C_6$ and a fully {\it ab
initio} curve for LiNa$+$LiNa.}
\end{figure}

In the low-field limit, coupling between rotational states can be limited to
just two states, allowing Eq.(\ref{rothamiltonian}) to be solved analytically
(note that $\Omega\equiv 0$ and $M=0$, as discussed below in Sec. \ref{ressec}).
>From this it is possible to obtain general expressions for the expectation value
of the static and alignment moments as a function of the applied field.
Transforming to the unitless field parameter $\xi=\xi_0 F$, with $\xi_0={\cal
D}/2B$, the low-field limit is defined by $\xi\leq 1$.  With this
approximation, the dressed state dipole and quadrupole moments can be shown to
be 
\begin{equation}\label{lowq10}
\langle Q^{\rm DS}_{10}\rangle(\xi) = \langle Q_{10}\rangle
\frac{6\xi+2\xi^3} {8(1+\xi^2)} \;,
\end{equation}
and
\begin{equation}\label{lowq20}
\langle Q^{\rm DS}_{20}\rangle(\xi) = \langle Q_{20}\rangle
\frac{2\xi^2} {15(1+\xi^2)} \;,
\end{equation}
respectively, while the octupole moment has no two state contribution by symmetry.
The orientation moment $\langle\cos\theta\rangle$ is given trivially by 
\begin{equation}\label{lowcos}
\langle\cos\theta\rangle(\xi) = \langle Q^{\rm DS}_{10}\rangle(\xi)/\langle Q_{10}\rangle,
\end{equation}
while alignment $\langle\cos^2\theta\rangle$ can be calculated by noting
that $\cos^2\theta=\frac{1}{3}(1+2C_{1,0}(\theta))$ (where $C_{\ell , m}(\hat{r}_i)$ 
is a Racah spherical harmonic), providing the expression
\begin{equation}\label{lowcos2}
\langle\cos^2\theta\rangle(\xi) =
\frac{15+19 \xi^2}{45(1+\xi^2)}.
\end{equation}
So long as the number of coupled states is dominated by the first two states and
$\xi\leq1$, these approximate formula are accurate to a few percent.
In Table \ref{xi0table} we have evaluated $\xi_0$ for all the heteronuclear
alkali diatoms from the spectroscopic data in Table \ref{statictable}.
It is also possible to evaluate Eqs.(\ref{transvdw}) and (\ref{dens}) using the
two state low-field approximation.  Following the prescribed method discussed above,
the low field-dressed-state van der Waals potential can be written to leading
order as
\begin{eqnarray}
E^{\rm DS}_{\rm 2st}({\bf R},\xi) &\simeq &
\frac{\tilde{W}^{(1)}_{320}(\theta_F,\xi)}{R^3} 
+\frac{\tilde{W}^{(1)}_{540}(\theta_F,\xi)}{R^5} \nonumber \\ & &
-\frac{W^{(2)}_{6000}}{R^6}-\frac{W^{(2)}_{8000}}{R^8}.
\label{2ste}
\end{eqnarray}
Here $\theta_F$ is the angle between the inter-molecular vector ${\bf R}$ and
the field vector as illustrated in Fig. \ref{geomfig}(b).  The dipole-dipole and
quadrupole-quadrupole contributions are (up to order $\xi^5$)
\begin{eqnarray}
\tilde{W}^{(1)}_{320}(\theta_F,\xi) & = &
\langle Q_{10}\rangle^2
\frac{3\sqrt{3}\xi+6\xi^2+4\sqrt{3}\xi^3+4\xi^4}
{27(1+\xi^2)^2} \nonumber \\ & & \times
(1-3\cos^2\theta_F)\;,
\end{eqnarray}
and 
\begin{eqnarray}
\tilde{W}^{(1)}_{540}(\theta_F,\xi) & = &
\langle Q_{20}\rangle^2
\frac{\xi^4}
{75(1+\xi^2)^2} \nonumber \\ & & \times
(3 - 30\cos^2\theta_F + 35\cos^4\theta_F) \;,
\end{eqnarray}
respectively (note that there is no dipole-octupole contribution in the two
state approximation) while $W^{(2)}_{n000}$ is the isotropic
dispersion+induction coefficient (see Tables \ref{livdwtable} and
\ref{othervdwtable}). The anisotropic terms contribute less than a percent to the
interaction energy and can be safely neglected.

\section{\label{abinitio}Electronic structure calculations}

\begin{figure*}
\resizebox{17cm}{!}{\includegraphics{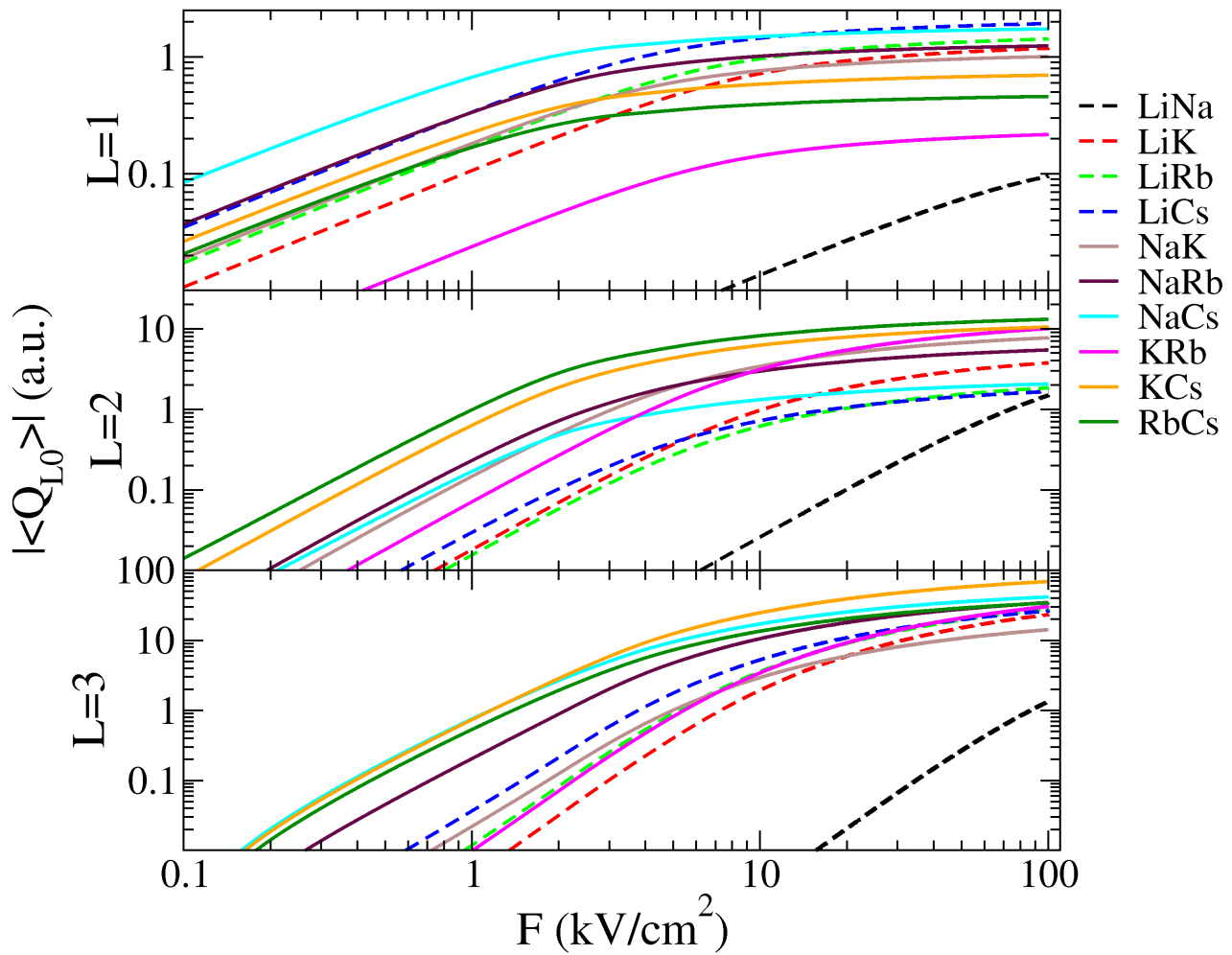}}
\caption{\label{elstfig}Dressed state electrostatic moments, $\langle
Q^{DS}_{\ell 0}\rangle,$ of various heteronuclear alkali diatomic molecules as a
function of an external DC electric field.}
\end{figure*}

The {\it ab initio} calculation of van der Waals coefficients, and more
generally multipole polarizabilities, requires special care in both the basis
set and level of theory used \cite{urban1995,labello2005,labello2006}.
Electrostatic moments similarly require careful consideration of the theoretical
method, though the basis set dependence is less severe \cite{harrison2005}.  For
all calculations in this work we use the Karlsruhe def2-QZVPP \cite{weigend2003}
basis set augmented with two additional even tempered diffuse $spdf$ functions
designed to accurately describe higher order static polarizabilities
\cite{byrd2011}.  The Karlshruhe def2 basis sets are available for nearly the
entire Periodic Table, and are known for both their robustness and good
cost-to-performance ratio in large molecular Hartree-Fock and density functional
theory calculations.  As such they remain attractive for use in calculations
that involve many different atoms across the period table.  

As has been demonstrated previously, the use of time dependent density
functional theory \cite{tawada2004,dreuw2005} is a cost effective and accurate
way to calculate multipole transition moments and excitation energies for
diatomic molecules.  We chose to limit our calculations in this work to only
include the PBE0 functional for simplicity, however for various cases it was
observed that the B3PW91 functional also provides consistent results.  The
electrostatic moments were calculated using coupled cluster theory including all
singles, doubles and perturbative triples (CCSD(T)) \cite{pople1987} using a two
step finite field method (with field spacings of $10^{-6}$ a.u.).
Core-valence and core-core correlation energy was accounted for by including the
inner valence $s$ and $p$ electrons in the CCSD(T) calculations, while for the
TD-DFT computations all electrons not replaced by an ECP are implicitly
correlated.  All TD-DFT calculations were done using a locally modified version
of the GAMESS \cite{gamess1993,gamess2005} suite of programs; the CCSD(T) finite
field calculations were done using the MOLPRO \cite{molpro10} quantum chemistry
program package.  For further details on the methodology used in evaluating the
transition dipole moments and excitation energy we refer to our previous paper
on homonuclear alkali diatomic molecules \cite{byrd2011}.

\section{\label{ressec}Computational results and discussion}

\begin{figure*}[t]
\resizebox{14cm}{!}{\includegraphics{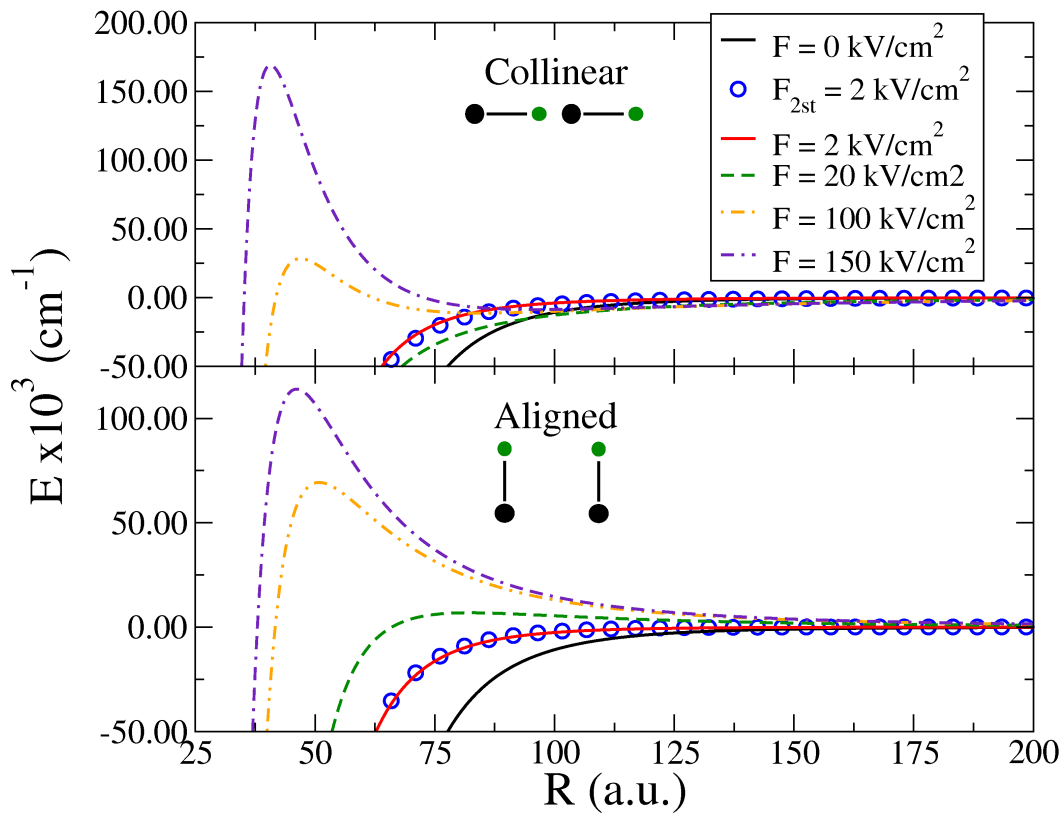}}
\caption{\label{dsvdwfig}
DC field coupled van der Waals curves (Eq.
(\ref{transvdw})) of $^{40}$K$^{87}$Rb for both low and high fields as well as
the approximate two-state van der Waals curve (Eq. (\ref{2ste})).
Here $F\sim 20$ kV/cm is the intermediate field strength where more than two
rotational states begin to strongly couple.}
\end{figure*}

The leading order term of the long range expansion 
Eq.(\ref{wsimple}), and thus the longest ranged
interaction in the series, involve products of the electrostatic moments 
of each monomer, and for dipolar molecules, it is the dipole-dipole $R^{-3}$ term.  
The dipole-dipole scattering \cite{julienne2011} and applications of
dipole-dipole interactions
\cite{yelin2009} are well studied in the literature, however higher order terms
can be necessary for accurately describing intermediate intermolecular distances
\cite{byrd2011} and are often neglected if only for a lack of available data.
Inclusion of just the quadrupole-quadrupole interaction to a dipole-dipole model
can introduce significant changes in the form of potential energy barriers for
collinear geometries ($\theta_1=\theta_2=\phi=0$) at long range
\cite{byrd2012-c}.
It is possible to estimate whether the inclusion of higher order electrostatic
terms could lead to a barrier by introducing the outer zero energy turning
point, $R_q,$ which occurs when the $R^{-5}$ repulsion overcomes the attractive
$R^{-3}$ dipole-dipole force. Keeping only the two leading terms in
Eq.(\ref{wsimple}) and setting 
\begin{equation}
E_{\rm int}(R_q,0,0,0) = 
-\frac{\langle Q_{10}\rangle^2}{R_q^{3}} 
+\frac{3\langle Q_{20}\rangle^2-4\langle Q_{10}\rangle\langle Q_{30}\rangle}
{R_q^{5}} = 0, 
\end{equation}
we obtain
\begin{equation}
R_q = 
\frac{\sqrt{3\langle Q_{20}\rangle^2-4\langle Q_{10}\rangle\langle Q_{30}\rangle}}
{\langle Q_{10}\rangle}.
\end{equation}
When this outer turning point is sufficiently long range ($R_q\gtrsim 20$ a.u.) the
introduction of these higher order terms can be important and lead
to long range barriers \cite{byrd2012-c}, and thus should be examined
in further detail.  As such, we have calculated the {\it ab initio}
electrostatic dipole, quadrupole and octupole moments (higher order moments do not
contribute up to $R^{-5}$ in the long range expansion: see Sec.
\ref{abinitio} for details on the methodology used).  In Table \ref{statictable}
we present our calculated static moments, the outer turning point $R_q$ for each
system, as well as various dipole and quadrupole moments found in the
literature.  Our computed static dipole moments agree closely with both the
valence full configuration interaction results of Aymar {\it et al.}
\cite{aymar2005} across all the molecules investigated, and the CCSDT (CCSD with
all triples) results of Qu\'em\'ener {\it et al.} \cite{quemener2011} for the
highly polar LiX (X$=$Na,K,Rb,Cs) species.  Other than the CCSD(T) quadrupole
moment of Zemke {\it et al.} \cite{zemke2010} (with which we compare well),
little to no published quadrupole values exist for the heteronuclear alkali
diatoms.  It has been demonstrated for the homonuclear alkali diatoms that 
the finite field CCSD(T) higher order static moments compare well with other methods
\cite{harrison2005,byrd2011}; similar accuracy is anticipated for the
heteronuclear species.

Dispersion and induction contributions to the van der Waals series are
proportional to products of the dipole, quadrupole and octupole
polarizabilities.  As such we have calculated and presented in Table
\ref{polartable} the dipole and quadrupole static polarizabilities with
comparisons to some of the existing literature (octupole static
polarizabilities are not listed, but are available upon request).  As discussed
previously \cite{byrd2011}, the n-aug-def2-QZVPP basis sets are well
converged for computation of static polarizabilities of homonuclear alkali
diatoms up to octupole order, and we find the same is true for the heteronuclear
species.  Density functional methods are known to provide average static
polarizabilities to within five to ten percent of experimental or highly correlated
results \cite{calaminici1998,adamo1999}.  Furthermore some variance is expected
in the parallel ($\alpha_{110}$) polarizability as all computations are done at
the experimental (or theoretical where necessary) equilibrium bond length, and
it is well known that the polarizability is sensitive to the internuclear
separation in the alkali diatoms \cite{deiglmayr2008}.  It is expected that the
perpendicular polarizability ($\alpha_{111}$) should agree much more closely
with other methods, which we find to be the case as illustrated in Table
\ref{polartable}.

Van der Waals dispersion and induction coefficients of the heteronuclear alkali
diatoms are sparsely given in the literature.  Currently only a few values exist
and are restricted to isotropic contributions (corresponding to $W^{2,{\rm
DIS}}_{6000}$). The only systematically calculations are for the LiX species
\cite{quemener2011}.  In Table \ref{polartable} we note the reasonable agreement
between our reported TD-DFT isotropic $C_6=W^{2,{\rm DIS}}_{6000}$ dispersion
coefficients and the Tang-Slater-Kirkwood \cite{tang1969} values from
Qu\'em\'ener {\it et al.} \cite{quemener2011} for the LiX species.  Additionally
Kotochigova \cite{kotochigova2010} has calculated, using multi-reference
configuration interaction theory, the isotropic and anisotropic dispersion
coefficients of order $R^{-6}$ for both KRb and RbCs.  However, these values
contain non-Born-Oppenheimer contributions and so are not directly comparable to
our numbers; because of this we have not included these values in Table
\ref{polartable}.  To determine the accuracy of the van der Waals coefficients
calculated here, we have computed {\it ab initio} curves for LiNa$+$LiNa at two
different geometries using the
CCSD(T)-F12a/QZVPP (explicitly correlated CCSD(T)) level of theory
\cite{adler2007,knizia2009}.  These {\it ab initio} curves are plotted in Fig.
\ref{abinitfig} along with the electrostatic plus isotropic dispersion
approximation and the the van der Waals curves of this work including all
anisotropic terms through $R^{-8}$.  As can be seen, the TD-DFT van der Waals
curves agree to a few $cm^{-1}$ with the {\it ab initio} results \footnote{Even
at the highly correlated level of theory used in computing the {\it ab initio}
curves in Fig. \ref{abinitfig}, there is a several $cm^{-1}$ difference in
barrier heights between the F12a and F12b explicitly correlated methods
demonstrating the difficulty in obtaining reliable results for molecular
interaction barriers at long range.}, while the isotropic curves fail completely
in the intermediate-range (it should be noted that for the collinear case of
LiNa$+$LiNa the isotropic curves do not turn over at all and predict an infinite
repulsive wall).
In Tables \ref{livdwtable} and
\ref{othervdwtable} we have listed the $W^{(1,2)}_{n L_1 L_2 M}$ coefficients
for all of the heteronuclear alkali diatoms, including all terms up through
order $R^{-8}$.  

In evaluating the field coupling and alignment of the various alkali diatomic
molecules, the rotational wavefunction expansion is greatly simplified by making
use of the initial premise that the molecules are in the ro-vibrational
ground state.  As such $\Omega\equiv 0$ and $M=0$ (the use of a DC external field
will not mix different $M$ values), reducing both Eqs.(\ref{transvdw}) and
(\ref{dressedq}) significantly.  In Fig. \ref{elstfig} we have plotted the DC
field dressed electrostatic moments as a function of the external field
strength.  While the very high field strengths in Fig. \ref{elstfig} are
generally experimentally challenging, it is illustrative to show how difficult
it is to obtain both strong orientation ($\langle\cos\theta\rangle>0.85$) and
alignment ($\langle\cos^2\theta\rangle>0.85$) in molecules with small rotational
constants, regardless of the strength of the dipole moment.  It is also
instructive to examine the low-field strengths of Fig. \ref{elstfig}, where the
linear trend of each curve on the log-log scale shows the general scaling of
the static moments as a function of the external field as discussed in Sec.
\ref{lowfs}.
In Fig.\ref{dsvdwfig} we have evaluated Eq.(\ref{transvdw}) for KRb (KRb is chosen for
its medium strength dipole moment and large rotational constant) at various DC
field strengths.  The difference between the low and high field strengths is
easily identified by the change in behavior from most similar to the field free
case (e.g. isotropic contributions dominate the interaction potential) to the
regime where the dressed state van der Waals interaction energy more closely
resembles the molecule-fixed frame van der Waals potential (e.g. when
electrostatic contributions become key).  This high field strength regime is
more quantitatively defined when both $\langle\cos\theta\rangle$ and
$\langle\cos^2\theta\rangle$ is greater than $0.9$ (which corresponds to roughly
$7$ strongly coupled rotational states).  
Also in Fig.\ref{dsvdwfig} the
approximate two-state model of Eq.(\ref{2ste}) can be seen to agree very well
with the fully coupled equations in the low-field limit.
Fully field-coupled potentials for the other heteronuclear molecules listed in this
work have been calculated, and are available upon request.

\begin{table}[t]
\begin{ruledtabular}
\caption{\label{xi0table}Tabulated values of the field strength coefficient
$\xi_0={\cal D}/2B$ using the spectroscopic and electrostatic constants from Table
\ref{statictable}.  All units are in cm$^2/$kV.}
\begin{tabular}{ldddd}
    & \multicolumn{1}{c}{$^{23}$Na}
    & \multicolumn{1}{c}{$^{39}$K}
    & \multicolumn{1}{c}{$^{95}$Rb}
    & \multicolumn{1}{c}{$^{133}$Cs}\\
\hline
 $^7$Li    & 0.0114 & 0.116 & 0.159 & 0.246 \\
 $^{23}$Na &  & 0.253 & 0.413 & 0.684 \\
 $^{39}$K  &  &  & 0.141 & 0.529 \\
 $^{95}$Rb &  &  &  & 0.635 
\end{tabular}
\end{ruledtabular}
\end{table}

\section{conclusions}

This work completes our systematic TD-DFT computation of the alkali diatomic
species by computing accurate multipole electrostatic moments and anisotropic
van der Waals coefficients for the heteronuclear alkali diatomic species.  The
multipole electrostatic moments were computed using a finite field treatment of
the CCSD(T) molecular energy employing the augmented Karlsruhe def2-QZVPP basis
set and found to produce excellent agreement with the existing literature.
Excitation energies and multipole transition moments were calculated using
TD-DFT and the same augmented QZVPP basis set.  Static polarizabilities as well
as van der Waals induction and dispersion coefficients were evaluated using the
sum over states approach and found to be consistent with the existing
literature.  Using the simple form of Eq.(\ref{wsimple}) and the values from
Tables \ref{livdwtable} and \ref{othervdwtable}, it is possible to completely
characterize the long range interaction between two heteronuclear alkali diatoms
up through order $R^{-8}$.  A sample FORTRAN program for evaluating
Eq.(\ref{wsimple}) is included in the supplemental material of Ref.
\cite{byrd2011} or upon request to the authors.

The transformation of the van der Waals series for linear molecules from the
molecule-fixed frame to the lab-fixed frame was described.  This was followed
with the computation of the dressed state electrostatic moments as a function of
an external DC electric field.  It was noted that in the low field limit, the
coupling of the molecule to the external field can be approximated by only
considering two rotational states.  With this in mind, the orientation and
alignment of the molecule as a function of the applied field can be approximated
using only molecular spectroscopic constants by
Eqs.(\ref{lowq10})-(\ref{lowcos2}), which are valid for field values $F \leq 2
B/{\cal D}$.  We have also illustrated the effects of an external DC electric
field on the intermolecular potential by evaluating Eq.(\ref{transvdw}) for
$^{40}$K$^{87}$Rb at a variety of field strengths.  It can be seen then that
introducing rotational state coupling leads to a richer interaction phase space
beyond the usual isotropic approximations.  Finally a two-state approximation of
the dressed-state long range potential (see Eq.(\ref{transvdw})) has been
derived and given by Eq.(\ref{2ste}) in terms of molecular spectroscopic
constants and isotropic van der Waals coefficients

\section{Acknowledgments}

J.B. would like to thank the Department of Defense Air Force Office of
Scientific Research MURI grant for support, and R.C. the Chemical Science,
Geoscience and Bioscience Division of the Office of Basic Energy Science, Office
of Science, U.S. Department of Energy.

%
%


%

\end{document}